\documentclass[aps,onecolumn,prx,preprintnumbers,showpacs,amsmath,amssymb,superscriptaddress,11pt]{revtex4-1}
\usepackage[pdftex, colorlinks, citecolor=blue]{hyperref}   
\usepackage{graphicx}
\usepackage{dcolumn}
\usepackage{threeparttable}
\usepackage{multirow}
\usepackage{booktabs}
\usepackage{txfonts}
\usepackage{xcolor}
\usepackage{bm}
\usepackage{amssymb}
\usepackage{amsmath}
\usepackage{latexsym}
\usepackage{epsfig}
\usepackage{amsbsy}
\usepackage{array}
\usepackage{tabularx}
\usepackage{esvect} 
\usepackage{extarrows}
\usepackage{float}
\usepackage[british]{babel}
\usepackage{longtable}
\usepackage{makecell}

\begin{document}

\title{Hardness and fracture toughness models by symbolic regression}	

\author{Jinbin Zhao}
\affiliation{%
Shenyang National Laboratory for Materials Science, Institute of Metal Research, Chinese Academy of Sciences, 110016 Shenyang, China.
}
\affiliation{%
	Taiyuan University of Science and Technology, School of Materials Science and Engineering, 030024 Taiyuan, China.
}%

\author{Peitao Liu}%
\email{ptliu@imr.ac.cn}
\affiliation{%
Shenyang National Laboratory for Materials Science, Institute of Metal Research, Chinese Academy of Sciences, 110016 Shenyang, China.
}

\author{Jiantao Wang}
 \affiliation{School of Materials Science and Engineering, University of Science and
Technology of China, 110016, Shenyang, China}
\affiliation{%
Shenyang National Laboratory for Materials Science, Institute of Metal Research, Chinese Academy of Sciences, 110016 Shenyang, China.
}%

\author{Jiangxu Li}
\affiliation{%
Shenyang National Laboratory for Materials Science, Institute of Metal Research, Chinese Academy of Sciences, 110016 Shenyang, China.
}%

\author{Haiyang Niu}
 \affiliation{State Key Laboratory of Solidification Processing, International Center for Materials Discovery, School of Materials Science and Engineering, Northwestern
Polytechnical University, Xi'an 710072, P. R. China. }

\author{Yan Sun}
\affiliation{%
Shenyang National Laboratory for Materials Science, Institute of Metal Research, Chinese Academy of Sciences, 110016 Shenyang, China.
}%

\author{Junlin Li}
\affiliation{%
	Taiyuan University of Science and Technology, School of Materials Science and Engineering, 030024 Taiyuan, China.
}%

\author{Xing-Qiu Chen}%
 \email{xingqiu.chen@imr.ac.cn}
\affiliation{%
Shenyang National Laboratory for Materials Science, Institute of Metal Research, Chinese Academy of Sciences, 110016 Shenyang, China.
}%

\begin{abstract}
Superhard materials with good fracture toughness have found wide industrial applications,
which necessitates the development of accurate hardness and fracture toughness models for efficient materials design.
Although several macroscopic models have been proposed, they are mostly semiempirical based on prior knowledge or assumptions,
and obtained by fitting limited experimental data.
Here, through an unbiased and explanatory symbolic regression technique,
we built a macroscopic hardness model and fracture toughness model,
which only require shear and bulk moduli as inputs.
The developed hardness model was trained on an extended dataset,
which not only includes cubic systems, but also contains non-cubic systems  with anisotropic elastic properties.
The obtained models turned out to be simple, accurate, and transferable.
Moreover, we assessed the performance of three popular deep learning models
for predicting bulk and shear moduli, and
found that the crystal graph convolutional neural network
and crystal explainable property predictor perform almost equally well,
both better than the atomistic line graph neural network.
By combining the machine-learned bulk and shear moduli
with the hardness and fracture toughness prediction models,
potential superhard materials with good fracture toughness
can be efficiently screened out through high-throughput calculations.
\end{abstract}

\maketitle

\section{Overview of hardness and fracture toughness models}

\subsection{Superhard materials}

Superhard materials have been widely used in many industrial applications due to their unique mechanical properties.
They not only have high hardness (usually greater than 40 GPa)~\cite{Haines2001SynthesisAD},
but also often have high fracture toughness and thermal stability~\cite{Xu2015SuperhardMR,Gilman2006DesignOH,Ivanovskii2011TheSF,Riedel1994NovelUM}.
The hardest known material is diamond, which has a Vickers hardness in the range of 60$\sim$120 GPa.
In addition, novel metastable carbon allotropes with comparable Vickers hardness
have been experimentally synthesized~\cite{Mao2003BondingCI} or theoretically predicted~\cite{Li2009SuperhardMP,PhysRevLett.108.135501,Avery2019PredictingSM}.
Some other potentially hard and superhard materials include
borides~\cite{doi:10.1063/1.4826485,Dong2018BoronOU,Zhao2018UnexpectedSP,D2TA02268K},
carbides~\cite{Chuvildeev2015HighstrengthUT,Sung1997ReactivitiesOT},
and nitrides~\cite{Gao2020SimultaneousEO,Ivashchenko2015FirstprinciplesQM}.
Among these materials, cubic boron nitride achieves a second largest hardness of $\sim$60 GPa~\cite{Tian2013UltrahardNC,Li2015HighEnergyDA}.
In particular, the superhard materials at the nanoscale have achieved excellent mechanical performances~\cite{Zhao2016RecentAI,Hovsepian2006SynthesisSA}.
This encourages ongoing exploration of reliable models for accurate prediction of hardness and fracture toughness.
This is particularly relevant in the context of data-driven high-throughput
screening and predictions~\cite{PhysRevB.98.014107,doi:10.1063/1.5109782,Mazhnik2020ApplicationOM,Avery2019PredictingSM}.

\subsection{Hardness and hardness prediction models}

Hardness has long been used as one of the fundamental mechanical properties of materials since 1772~\cite{cole1957}.
However, it is not as well defined as other mechanical properties like strength and plasticity~\cite{Haines2001SynthesisAD}.
Macroscopically, the hardness of a material is defined as its ability to resist being scratched or dented by other materials.
In experiments, the indentation machine is normally used to measure the hardness.
According to the different shapes and properties of the indenters,
several flavors of the hardness have been developed, e.g.,
the Brinell hardness, Rockwell hardness, Knoop hardness, and Vickers hardness.
Among them, the Vickers and Knoop hardness are most commonly used in standardized tests for engineering and metallurgy~\cite{NIX1998411}.

Experimentally, hardness is a highly complex property, as the stress applied depends on many factors
such as the crystallographic direction, loading force, and the size of the indenter.
The same holds also for theory, because the microscopic understanding of the hardness
is far from complete, which makes an accurate prediction of the hardness difficult~\cite{SUN2022215}.
Despite of this, many microscopic and macroscopic hardness prediction models have been established,
with their profound success in the search and prediction of new superhard materials~\cite{Li2010PredictingNS,TIAN201293}.

By definition, the hardness of a material is intimately linked to its elastic properties.
As early as in 1973, Gilman obtained a linear relationship between the hardness and bulk modulus ($B$)~\cite{Gilman1973HardnessOP}.
Then, in 1998 Teter identified a strong correlation between the hardness and shear modulus
$H^{\rm Teter}_V = 0.151 G$~\cite{teter_1998}.
Subsequent study, however, revealed that the hardness is not simply linear to the bulk modulus or shear modulus~\cite{CHEN20111275}.
Since the Pugh's modulus ratio $k=G/B$~\cite{doi:10.1080/14786440808520496}
is closely related to the brittleness/ductility of the material and
underlines the relationship between the plastic and elastic properties of pure polycrystalline metals~\cite{PMID:23056910,Senkov2021GeneralizationOI},
Chen \emph{et al.}~\cite{CHEN20111275} proposed a macroscopic hardness model by incorporating the Pugh's modulus ratio
\begin{equation}\label{eq:Chen}
	H^{\rm Chen}_V = 2(k^2G)^{0.585}-3.
\end{equation}
The Chen's model has been well appreciated by the community in
predicting the hardness of various crystalline materials~\cite{Chen2011HardnessOT,Zhang2014FeB4,ZHANG2017802,D2RA01593E},
because the model is very simple,
only requires two input parameters ($G$ and $B$) that
can be routinely calculated through first-principles calculations,
and more importantly, reproduces the experimental hardness data remarkably well~\cite{CHEN20111275}.
Nevertheless, the Chen's model was obtained by fitting the experimental hardness data
combined with semiempirical domain knowledge.
The data used in fitting are mostly cubic crystals with isotropic elastic properties,
which may pose potential risks to the application of the model to systems with strong anisotropic elastic properties.
Besides, Tian \emph{et al.}~\cite{TIAN201293} noticed that
the Chen's model may yield unphysical negative values when predicting the materials with small hardness
due to the presence of an intercept term ($-3$) in the model. To avoid this,
they refitted the experimental data compiled in the Chen's work
and proposed a modified formula without the intercept term
$H^{\rm Tian}_V = 0.92k^{1.137}G^{0.708}$~\cite{TIAN201293}.
As expected, both Chen's and Tian's models reproduce well the experimental values that were used for fitting.
However, when predicting low-hardness materials (e.g., less than 5 GPa), both formulas tend to overestimate the predictions.
Motivated by the Chen's model, Mazhnik and Oganov rewrote the model in terms of Young's modulus $E$ and Poisson's ratio $\nu$
using the homogeneous approximation and established a relationship between the hardness and the effective modulus~\cite{Mazhnik2019AMO}.
The effective modulus was obtained by fitting the experimental data using a polynomial rational function~\cite{Mazhnik2019AMO}.
Although the polynomial rational function increases the flexibility of the model, the resulting hardness model loses the simplicity.
In addition, the data used for fitting were also limited to the dataset collected in the Chen's work, most of which are cubic systems.

In contrast to the macroscopic hardness models,
the microscopic hardness models reply on the information (e.g., bond density, bond length, and bonding type, etc.)
that can either be derived directly from the crystal structure or extrapolated from the
constituent elements~\cite{Gao2003HardnessOC,Guo2008HardnessOC,Li2008ElectronegativityIO,imnek2007HowTE,PhysRevLett.96.085501,TIAN201293,Dai2016AMT}.
These include the bond resistance models (Gao model~\cite{Gao2003HardnessOC} and Guo model~\cite{Guo2008HardnessOC}),
electronegativity model (Xue model)~\cite{Li2008ElectronegativityIO} and bond strength model (SV model)~\cite{PhysRevLett.96.085501}.
These microscopic hardness models have been successfully applied to covalent and polar covalent crystals, and in some cases, to ionic crystals~\cite{TIAN201293}.
Although the microscopic models have their own merits, they are normally not as simple as those macroscopic models.
Furthermore, the microscopic models may suffer from the underlying assumptions and predict incorrect hardness values.
A typical example is the so-called T-carbon, a novel hypothetical carbon allotrope~\cite{Sheng2011TcarbonAN}.
It has a porous structure with a highly anisotropic distribution of the $sp^3$-like C-C bonds,
and thus is unlikely to be a superhard material.
However, the Gao~\cite{Gao2003HardnessOC} and SV~\cite{PhysRevLett.96.085501} microscopic models
incorrectly predicted the T-carbon to be a superhard material, with an overestimated hardness of 61.1 GPa and 40.5 GPa, respectively.
By contrast, the Chen's macroscopic model yielded a reasonable value of hardness (5.6 GPa) for the T-carbon~\cite{Chen2011HardnessOT}.
The success of the macroscopic hardness models benefits from the fact that they do not explicitly reply on
those not so well-defined quantities like bond electronegativity or bond strength.
It is worth noting that recently Podryabinkin \emph{et al.}~\cite{doi:10.1021/acs.jctc.1c00783} developed a method
that allows to calculate the nanohardness by atomistic simulations of nanoindentation, 
benefiting from efficient machine-learning interatomic potentials.
Despite being accurate, the method is computationally too demanding to enable high-throughput calculations.

\subsection{Fracture toughness and its prediction models}

Apart from the hardness, searching for hard materials with good fracture toughness is equally important.
This is particularly important for practical applications, since the fracture toughness
describes the resistance ability of a material against crack propagation
and is often an indication of the amount of stress required to propagate a preexisting flaw.
The fracture toughness can be quantitatively described by the stress intensity factor $K$,
 at which a thin crack in the material starts to grow~\cite{b3003450e3cb4cceb0e4cfeab32e40bf}.
Similarly to the hardness, a large uncertainty is also present in the experimental fracture toughness data,
since the values of the fracture toughness are largely  influenced by the experimental details like loadings
and underlying fracture propagation mechanism.
This necessitates the development of fracture toughness prediction models.

However, unlike the hardness models, relatively less efforts have been devoted to developing fracture toughness prediction models.
The critical value of the stress intensity factor under mode I loading~\cite{doi:10.1098/rsta.1921.0006} is given by
$K_g=2\sqrt{\gamma_sG/(1-\nu)}$~\cite{THOMSON1987965}, where $\gamma_s$ is the surface energy of the material and $\nu$ is the Poisson’s ratio.
This was derived by balancing the surface tension ($2\gamma_s$) of the opening surfaces at the crack tip
and the elastic driving force.
In practice, $K_g$ is much higher than the actual measured value $K_{IC}$,
and thus represents the upper bound of fracture toughness~\cite{THOMSON1987965}.
Through studying the correlation between the fracture toughness and the elastic properties of materials,
Niu {\it et al.}~\cite{Niu_2019} developed a fracture toughness prediction model that works well for covalent and ionic crystals
\begin{equation}\label{eq:Niu}
K^{\rm Niu}_{IC}=V_0^{1/6}G(B/G)^{1/2},
\end{equation}
where $V_0$ is the volume of the system.
It is evident that $K^{\rm Niu}_{IC}$ is also related to the Pugh's modulus ratio $k=G/B$.
We note in passing that
Mazhnik and Oganov~\cite{Mazhnik2019AMO} rewrote the Niu's model in terms of Young's modulus and Poisson's ratio,
and proposed a new fracture toughness model based on the effective modulus,
which was again obtained by fitting the experimental fracture toughness data using a polynomial rational function~\cite{Mazhnik2019AMO}.

\subsection{Data-driven superhard materials predictions}

Due to the rapid increase of computing resources,
many large public datasets are available~\cite{Rajan2015MaterialsIT,Callister2015MaterialsSA}.
Most of the datasets are based on density functional theory (DFT) calculations,
e.g.,
Automatic  Flow  of  Materials  Discovery Library  (AFLOWLIB)~\cite{AFLOWLIB2012},
Open Quantum Materials Database  (OQMD)~\cite{OQMD2015},
Materials  Project  (MP)~\cite{articleMP},
and  Joint Automated   Repository   for   Various   Integrated   Simulations (JARVIS)~\cite{Choudhary2020}.
This allows us to efficiently search, predict and design materials with target properties,
and explore the underlying patterns through the data-driven
machine learning (ML) techniques~\cite{articleChoudhary,Morgan2020OpportunitiesAC,articlescience,
Morgan2020OpportunitiesAC,Sparks2020MachineLF,Suh2020EvolvingTM,Hart2021MachineLF,Saal2020MachineLI}.
The workflow of machine learning in general involves the collection and organization of training datasets,
design of material descriptors~\cite{Isayev2016UniversalFD,Alizadeh2019PredictingEC},
model regressions, and model validations and predictions.
The successful ML models include random forest (RF)~\cite{Stanev2017MachineLM,Ward2016AGM},
neural networks (NN)~\cite{Shi2019DeepES,Orponen1994ComputationalCO}, convolutional neural networks (CNN)~\cite{Tsymbalov2021MachineLF},
graph convolutional neural networks (GNN)~\cite{Chen2018GraphNA,Chen2021AtomSetsAA,Xie2017CrystalGC}, etc.
In the specific field of superhard materials,
Avery {\it et al.}~\cite{Avery2019PredictingSM} and Mazhnik {\it et al.}~\cite{Mazhnik2020ApplicationOM}
demonstrated the power of combining the ML-derived bulk and shear moduli and the macroscopic Vickers hardness models
in predicting novel superhard materials.
Tehrani {\it et al.}~\cite{MansouriTehrani2018MachineLD} showed how the elastic moduli predicted
by the support vector machine were successfully  used to guide the synthesis of superhard materials.

\subsection{The contribution of this work}

In this work, we attempt at building macroscopic hardness and fracture toughness models
through the unbiased symbolic regression technique
that does not need to make prior assumptions about the specific function form.
The dataset used in the regression extends the previously collected experimental dataset
(mostly consisting of cubic systems) and involves systems with anisotropic elastic properties.
The resulting hardness and fracture toughness models turn out to be very simple and
more accurate as compared to the Chen's hardness model and Niu's fracture toughness model.
Interestingly, the Pugh's modulus ratio $k=G/B$,
a quantity that is closely related to the brittleness/ductility of the material,
is naturally included in the regression models.
Moreover, we developed machine learning models for
efficient and accurate prediction of bulk and shear moduli,
the only two required input quantities for the hardness and fracture toughness models.
We compared and validated three popular deep learning algorithms
including the convolutional neural network,
atomistic line graph neural network, and crystal explainable property predictor.
With the bulk and shear moduli feeding to the hardness and fracture toughness regression models,
several potential superhard materials with good fracture toughness have been predicted through high-throughput calculations.

\section{Symbolic regression}\label{sec:level2}

\begin{figure}
	\centering
	\includegraphics[width=0.75\textwidth]{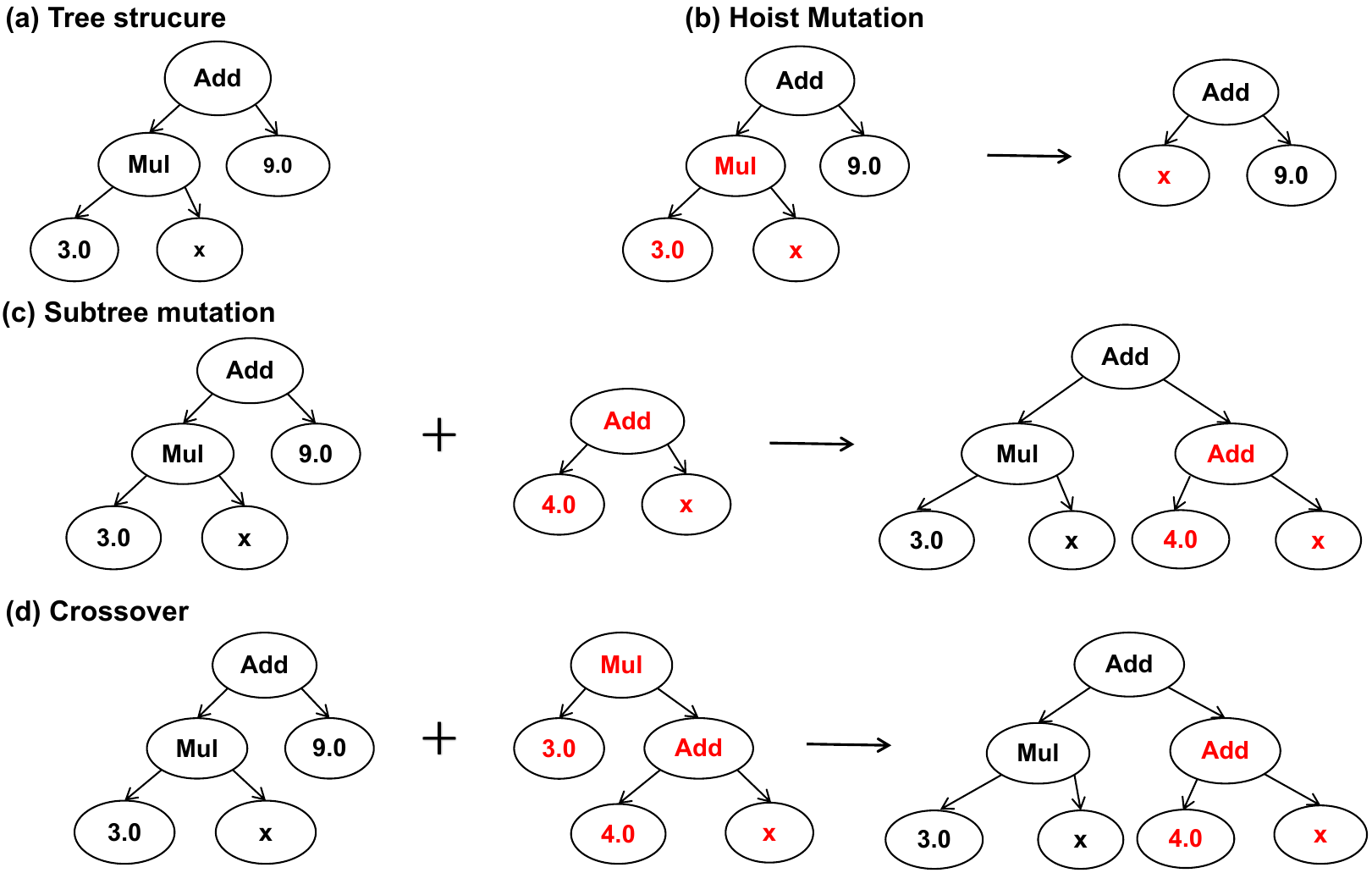}
	\caption{(a) Tree-structure chromosome representation of computer
programs in genetic programming.
(b) Hoist Mutation.
(c) Subtree mutation.
(d) Crossover.}
\label{Fig1_regression}
\end{figure}

Although ML models can be pretty predictable, they often contain thousands of parameters and thus require large datasets.
More importantly, the ML models typically lack of interpretability, which prevents them from searching for physical laws or finding a neat formula.
By contrast, symbolic regression (SR) is an interpretable machine learning algorithm.
As a supervised learning method, SR attempts to discover some hidden mathematical formulas
so as to predict target variables by using characteristic variables.
It is an effective regression method to search for the optimal form of a given set of functions and parameters,
even with limited data~\cite{wang_wagner_rondinelli_2019,WengSong-525,Stephens}.
For SR, there is no need to presuppose the specific composition of the functions.
Instead, one only needs to provide an expression that contains mathematical modules, such as mathematical operators, state variables, analytic functions, etc.
Then, SR explores the combinatorial  space using these constituent modules to obtain the most suitable scheme~\cite{wang_wagner_rondinelli_2019}.
The most common method of SR is genetic programming,
which was developed by Koza~\cite{Koza1994GeneticPA} as a concrete implementation of genetic algorithms (GA)~\cite{doi:10.1126/science.8346439}.
Inspired by the Darwin's theory of evolution, GA first proposes a tentative solution to a given problem,
and then finds the optimal solution after several iterations through crossover and mutation operations.
The solutions in genetic programming use a tree structure with nodes and terminals to represent chromosomes.
Figure~\ref{Fig1_regression}(a) shows a chromosome example of a mathematical function $3.0 \times x + 9.0$.
The tree consists of a set of interior nodes with mathematical operations [$+$ (Add) , $\times$ (Mul)]
and terminal nodes with variables ($x$) and constants (3.0, 9.0).
In order to obtain the final mathematical expression for each individual solution,
a depth-first search is performed for traversing the tree.

The process of symbolic regression starts from a set of randomly generated initial terminal nodes and functions to form trees with different sizes and structures.
In this work, the initial terminal nodes were the elastic moduli and randomly generated formula.
Then, the "fitness" of each individual solution
was evaluated by comparing the model predictions to the ground truth (i.e., experimental hardness or fracture toughness) in the dataset.
The fitness value is an assessment of the accuracy of the resulting formula.
The common error measures include mean absolute error (MAE) and root-mean squared error (RMSE).
The next generation was then evolved by performing random genetic manipulations
[e.g., mutations in Fig.~\ref{Fig1_regression}(b)-(c) and crossover in Fig.~\ref{Fig1_regression}(d)] on individual components.
The new generation was determined by following the "survival of the fittest" rule.
In this work, the symbolic regression was carried out using {\tt gplearn}~\cite{Stephens}, a Python library that
retains the familiar {\tt scikit-learn} fit/predict API and works with the existing {\tt scikit-learn} pipeline and grid search modules~\cite{Stephens}.
The hyper-parameters used in this work for {\tt gplearn} are given in Table~\ref{tab:symbolic},
and are briefly explained in the following.

For small datasets, a small population size and number of jobs (n-jobs) should be used,
which can accelerate the calculation and fast the convergence.
For large datasets, such as the extended hardness dataset,
the population size and n-jobs need to be appropriately increased in order to obtain a better accuracy.
The sum of the four parameters, the probability of crossover (pc),
subtree-mutation (ps), hoist-mutation (ph), point-mutation (pp) should be less than or equal to 1.
Among them, the ps is an important parameter for avoiding bloat,
a phenomenon often encountered in {\tt gplearn} where the program sizes grow larger and larger but with no significant improvement in fitness.
Besides the ps, another important parameter to fight against bloat is the parsimony coefficient: The larger the parsimony coefficient, the shorter the formula.

\setlength{\tabcolsep}{6.5pt}
\setlength{\LTcapwidth}{\textwidth}
\renewcommand\arraystretch{1.2}
\begin{table}
	\caption{The setup of hyper-parameters used in {\tt gplearn} for the symbolic regression.}
	\begin{tabular}{ccc}
		\hline
		\hline
		&Parameter	 & Value     \\
		\hline
		&population size	& 2000-8000 \\
		& n-jobs	& 3-20 \\
		& generations	& 50-200\\
		& stopping-criteria	& 0.01 \\
		& p-crossover (pc)	& 0.7(0.6) \\
		& p-subtree-mutation (ps)	& 0.2(0.25)\\
		& p-hoist-mutation (ph)	& 0.05\\
		& p-point-mutation (pp)	& 0.05 \\
		& p-point-replace (pr)   & 0.5 \\
		& parsimony coefficient	& 0.0001-0.01 \\
		& metric	& mean absolute error\\
		& function-set 	& add, sub, mul, div, sqrt, neg (mul, div, sqrt)  \\
		\hline
		\hline	
	\end{tabular}
\label{tab:symbolic}
\end{table}

\section{\label{sec:level3}{Results and discussions}}

\subsection{\label{sec:level1}{Modeling hardness of polycrystalline materials}}

\begin{figure}
	\centering
	\includegraphics[width=0.5\textwidth]{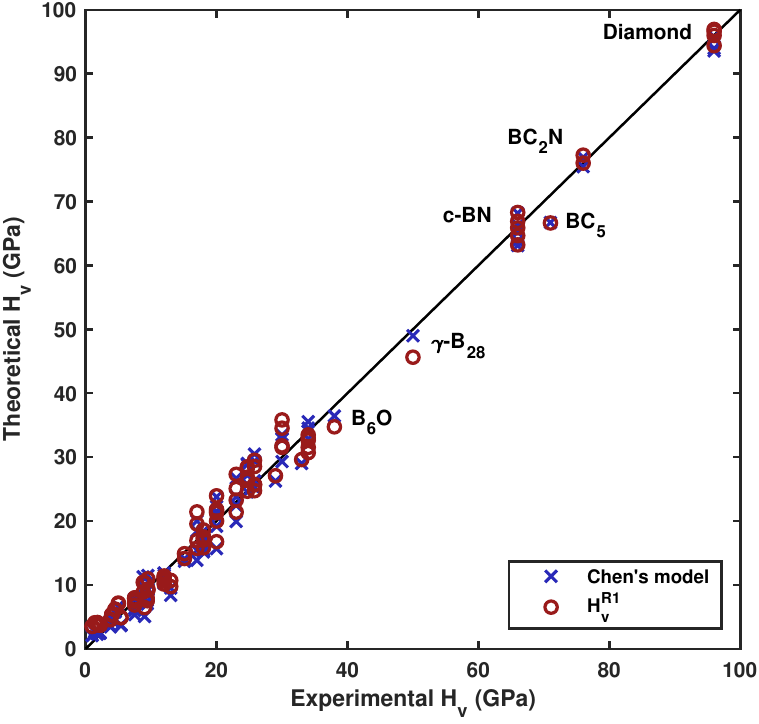}
	\caption{The predicted hardness versus the experimental data
using the Chen's model [Eq.~\eqref{eq:Chen}] and the model in Eq.~\eqref{eq:H_M1}.
Note that the assessment is performed on the dataset of $T^A$ (Table~\ref{tab:TA})
on which the two models were obtained.}
	\label{Fig2_model_I_train}
\end{figure}

Let us start from the symbolic regression on the dataset used in the Chen's work~\cite{CHEN20111275},
taking $B$ and $G$ as input descriptors.
We recall that the dataset (referred to as $T^{\rm A}$) contains mostly cubic
crystal structures with isotropic elastic properties.
The detailed information on $T^A$ is given in Table~\ref{tab:TA} of the Appendix.
The optimal harness model obtained from symbolic regression
is expressed as
\begin{equation}\label{eq:H_M1}
H^{\rm R1}_V=0.719kG^{3/4}.
\end{equation}
We mean that the model is optimal in the sense of
yielding the smallest MAE, but without at the cost of losing the model simplicity.
It is surprising to observe that, although $B$ and $G$ are the only two input descriptors,
the resulting hardness model naturally incorporates the Pugh's modulus ratio $k=G/B$.
This confirms that the Pugh's modulus ratio is indeed a good descriptor for the brittleness/ductility of a material.

Figure~\ref{Fig2_model_I_train} displayed the predicted hardness using the model in Eq.~\eqref{eq:H_M1} against the experimental data.
For comparison, the hardness predicted using the Chen's model [Eq.~\eqref{eq:Chen}] is also shown.
It can be seen that both models reproduce well the experimental data in $T^A$ with similar MAEs
($\sim$1.6 GPa) and RMSEs ($\sim$2.0 GPa) (see Table~\ref{tab:RMSE_H}).
We note that the Tian's model~\cite{TIAN201293} obtained by refitting $T^A$ exhibits similar performances (see Table~\ref{tab:RMSE_H}).

\begin{figure}
	\centering
	\includegraphics[width=0.95\textwidth]{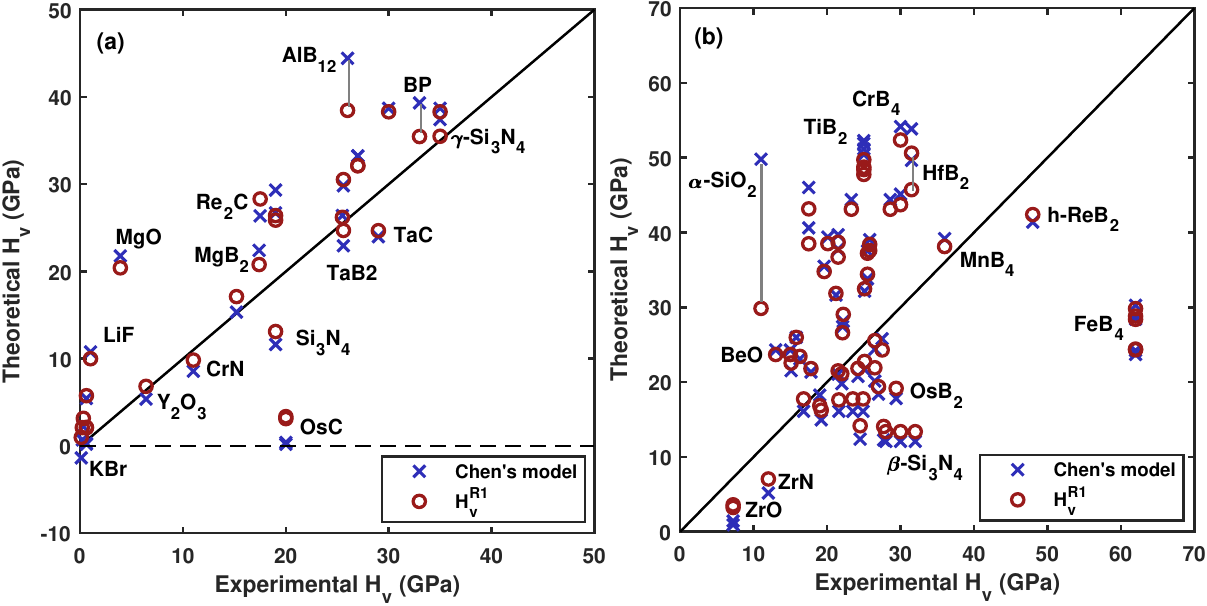}
	\caption{Assessment of the Chen's model  [Eq.~\eqref{eq:Chen}] and the model in Eq.~\eqref{eq:H_M1}
for the (a) dataset $T^B$ (Table~\ref{tab:TB}) and (b) dataset $T^C$  (Table~\ref{tab:TC}).}
\label{Fig3_model_I_test}
\end{figure}

\begin{table}
\renewcommand\arraystretch{1.1}
\caption{A summary of MAEs and RMSEs (in GPa) of various hardness models with respect to the experimental data.
The dataset used for the assessment is given in parentheses.
In the Oganov's model~\cite{Mazhnik2019AMO} $H_V = 0.096\chi(\nu)E$ where $\chi(\nu)=\frac{1-8.5\nu+19.5\nu^2}{1-7.5\nu+12.2\nu^2-19.6\nu^3}$
with $\nu$ and $E$ being the Poisson's ratio and Young's modulus, respectively. }
\resizebox{\textwidth}{3cm}{
	\begin{tabular}{c|cccccc}
		\hline
		\hline
	& \textbf{Teter}~\cite{teter_1998}& \textbf{Chen}~\cite{CHEN20111275}  & \textbf{Tian}~\cite{TIAN201293}  &
\textbf{Oganov}~\cite{Mazhnik2019AMO} &\textbf{Model $H^{\rm R1}_V$} [Eq.~\eqref{eq:H_M1}] &  \textbf{Model $H^{\rm R2}_V$}  [Eq.~\eqref{eq:H_M2}]  \\
	&$0.151G$	& $2(k^2G)^{0.585}-3$ & $0.92k^{1.137}G^{0.708}$ & $0.096\chi(\nu)E$ & $0.719kG^{3/4}$ & $0.16k^{1/2}G$\\
		\hline
		 MAE\,\,\,($T^A$)	&3.95& 1.58 &1.63 &4.74& 1.62&2.23 \\
		 RMSE ($T^A$)	&5.60& 2.01 &1.97 &7.44&2.03&3.05 \\
		\hline
		 MAE\,\,\,($T^B$)	&4.79& 6.35& 5.68&3.76& 5.47 &4.24\\
		 RMSE ($T^B$)	&6.59& 8.62 &7.61 &4.90&7.35 &6.11\\
       \hline
		 MAE\,\,\,($T^C$)	&10.31& 14.04 & 13.14&11.79&12.72&10.90 \\
		 RMSE ($T^C$)	&13.26& 17.42 & 16.40&15.78&15.91& 14.28\\
        \hline
         MAE\,\,\,($T^{\rm full}$)	&6.02& 6.27 & 5.89&5.72&5.70&5.31 \\
		 RMSE ($T^{\rm full}$)	&8.82& 10.30 & 9.65&9.26&9.37& 8.56\\
		\hline
		\hline	
	\end{tabular}
 }
\label{tab:RMSE_H}
\end{table}

In order to validate the hardness models, we generated two test datasets.
The first one contains the cubic systems that are not included in $T^A$ and is referred to as $T^B$.
The second one includes the non-cubic systems and is called  $T^C$.
The detailed information on $T^B$ and $T^C$ are provided in Table~\ref{tab:TB} and Table~\ref{tab:TC} of the Appendix, respectively.
The resulting assessment of the models is shown in Fig.~\ref{Fig3_model_I_test} with the
MAEs and RMSEs being given in Table~\ref{tab:RMSE_H}.
One can observe that the validation MAEs and RMSEs of the two models on $T^B$ and $T^C$
are significantly increased with the latter being more pronounced.
This is not unexpected, since the two models were obtained by fitting the data in $T^A$ containing most of cubic systems.
However, the model in Eq.~\eqref{eq:H_M1} exhibits an overall better performance
than the Chen's model (see Fig.~\ref{Fig3_model_I_test} and Table~\ref{tab:RMSE_H}).
This is manifested by the systems of cubic AlB$_{12}$, OsC and BP [Fig.~\ref{Fig3_model_I_test}(a)]
and trigonal $\alpha$-SiO$_2$  [Fig.~\ref{Fig3_model_I_test}(b)].
Furthermore, the Chen's model incorrectly predicts a negative hardness value for KBr due to the presence of intercept ($-$3) in the model,
whereas the model in Eq.~\eqref{eq:H_M1} restores the experimental value.
For the test datsets $T^B$ and $T^C$, the Tian's model~\cite{TIAN201293}  performs almost equally well with the model in Eq.~\eqref{eq:H_M1},
whereas the Teter's model~\cite{teter_1998} and Oganov's model~\cite{Mazhnik2019AMO} yield relatively better performances (see Table~\ref{tab:RMSE_H}).

\begin{figure}
	\centering
	\includegraphics[width=0.6\textwidth]{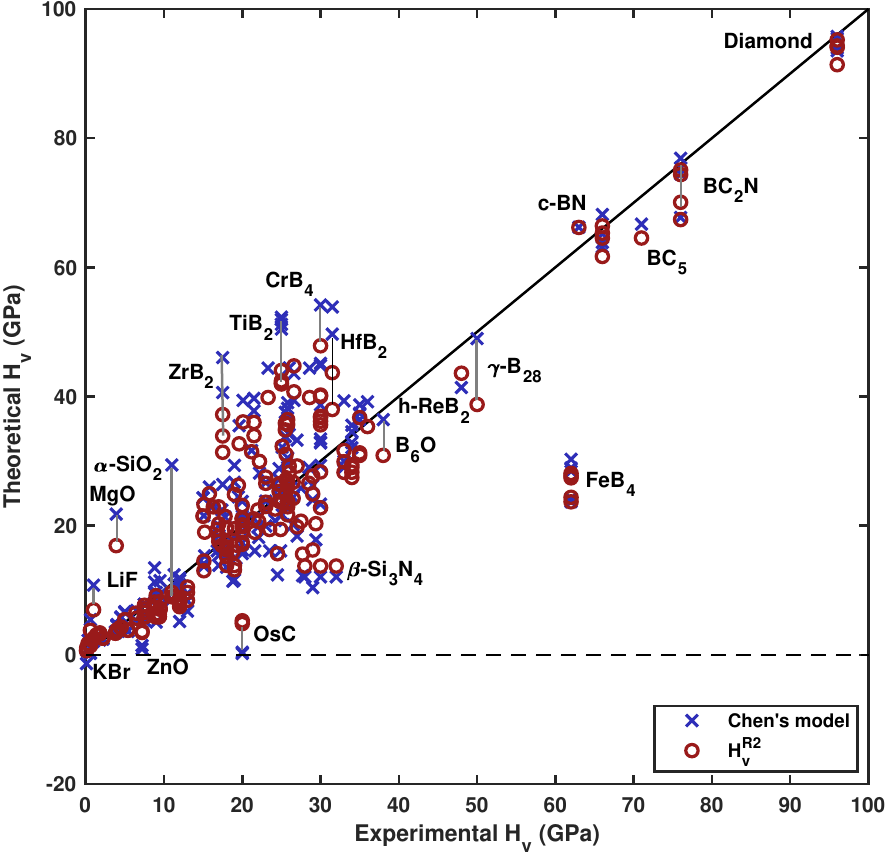}
	\caption{Assessment of the Chen's model [Eq.~\eqref{eq:Chen}] and the model in Eq.~\eqref{eq:H_M2} for the full dataset
including $T^A$, $T^B$ and $T^C$.}
\label{Fig4_model_II}
\end{figure}

To build a more broadly transferable hardness model, we combined all the data in $T^A$, $T^B$ and $T^C$
and randomly divided them into two parts with 90\% of the data for the symbolic regression and the remaining data for validating the model.
After repeating the regression several times, the optimal regression model in terms of accuracy and simplicity was selected
and expressed as
\begin{equation}\label{eq:H_M2}
H^{\rm R2}_V=0.16k^{1/2}G.
\end{equation}
Again, the Pugh's modulus ratio $k=G/B$ is naturally included by the symbolic regression.

Figure~\ref{Fig4_model_II} assesses the performance of the model in Eq.~\eqref{eq:H_M2} on the full dataset
and compares it to the Chen's model. The MAEs and RMSEs for each individual dataset are given in Table~\ref{tab:RMSE_H}.
It is evident that $H^{\rm R2}_V$ improves upon the Chen's model in the predictions for $T^B$ and $T^C$,
while the description for $T^A$ is somewhat slightly deteriorated.
This is, again, due to the fact that the Chen's model was fitted using only $T^A$.
One can also observe that the Chen's model predictions in fact are very accurate for materials in the superhard regime.
However, it severely overestimates the hardness values for materials in the intermediate-hard regime,
such as transition-metal borides like ZrB$_2$, TiB$_2$, HfB$_2$ and CrB$_4$.
For those materials with small hardness, the Chen's model tends to underestimate the experimental values
and even incorrectly predicts a negative hardness value for KBr.
Overall, the model in Eq.~\eqref{eq:H_M2} is more accurate,
yielding relatively smaller MAE and RMSE for the entire dataset as compared to the Chen's model.
Although the above direct comparison between $H^{\rm R2}_V$ and the Chen's model is not so fair,
our study highlights the importance of the training data in obtaining a more accurate and transferable model.
Furthermore, one can infer that modeling materials with anisotropic elastic properties is
more difficult as compared to the cubic systems with isotropic elastic properties.
This results in the dominant errors for all the hardness models (see Table~\ref{tab:RMSE_H}).
It is somewhat surprising that the simplest Teter's model~\cite{teter_1998}
also yields an overall good description of the entire dataset $T^{\rm full}$,
though it performs worse for the Chen's dataset $T^{\rm A}$.
This is mostly likely due to the strong correlation between $B$ and $G$,
as we will show later.

\subsection{\label{sec:level2}{Modeling fracture toughness of covalent and ionic crystals}}

Motivated by the success of symbolic regression in modeling the hardness,
we now turn to the modeling of fracture toughness using a similar procedure.
We employed the same fracture toughness dataset for covalent and ionic crystals
from Niu {\it et al.}~\cite{Niu_2019},
and took the shear modulus $G$, bulk modulus $B$ and atomic volume $V_0$ as input descriptors.
Several fracture toughness models were obtained and they were ranked according to the MAE.
The optimal fracture toughness model in the spirit of accuracy and simplicity reads
\begin{equation}\label{eq:K_regression}
K^{R}_{IC}=3.3 k^{1/8}B^{5/4}.
\end{equation}
Interestingly, this model only involves $G$ and $B$, whereas the atomic volume $V_0$
appearing in Eq.~\eqref{eq:Niu} is not present here.
This can be understood, because the fracture toughness is mainly connected to the elastic properties,
and the atomic volume  $V_0$  in Eq.~\eqref{eq:Niu} was simply used to yield the dimension of fracture toughness.

\begin{figure}
	\centering
	\includegraphics[width=0.6\textwidth]{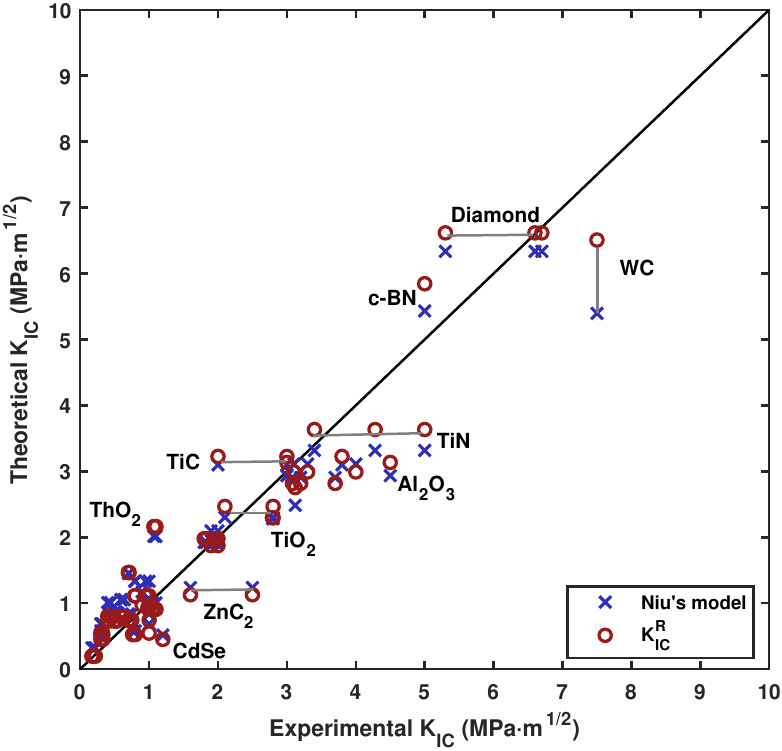}
	\caption{The predicted fracture toughness versus the experimental data
using the Niu's model [Eq.~\eqref{eq:Niu}] and the model in Eq.~\eqref{eq:K_regression}.
The experimental data are taken from Ref.~\cite{Niu_2019} and compiled also in Table~\ref{tab:fracture_toughness}.
}
\label{Fig5_toughness}
\end{figure}

\begin{table}
	\caption{A summary of MAEs and RMSEs  (in MPa$\cdot$m$^{1/2}$)  of fracture toughness models with respect to the experimental data.
In the Oganov's model~\cite{Mazhnik2019AMO} $\zeta(\nu)=\frac{1-13.7\nu+48.6\nu^2}{1-15.2\nu+70.2\nu^2-81.5\nu^3}$.}
	\begin{tabular}{cl|cc}
		\hline
		\hline
		&Models	& MAE & RMSE \\
		\hline
		& $K^{\rm Niu}_{IC}=V_0^{1/6}G(B/G)^{1/2}$ \cite{Niu_2019} & 0.44 & 0.60  \\
        & $K^{\rm Oganov}_{IC}=8840^{-1/2}V_0^{1/6}[\zeta(\nu)E]^{3/2}$ \cite{Mazhnik2019AMO}  &0.35&0.50\\
        & $K^{R}_{IC}=3.3 k^{1/8}B^{5/4}$    [Eq.~\eqref{eq:K_regression}]             & 0.37 & 0.54\\
		\hline
		\hline	
	\end{tabular}
\label{tab:RMSE_toughness}
\end{table}

Table~\ref{tab:fracture_toughness} of the Appendix collects the experimental fracture toughness data,
on which the Niu's model [Eq.~\eqref{eq:Niu}] and the model in Eq.~\eqref{eq:K_regression} were obtained.
The assessment of the two models against the experimental data is shown in Fig.~\ref{Fig5_toughness}.
The MAEs and RMSEs of the two models
as well as the Oganov's fracture toughness model~\cite{Mazhnik2019AMO}
are listed in Table~\ref{tab:RMSE_toughness}.
It can be seen that all three models reproduce well the experimental data.
The model in Eq.~\eqref{eq:K_regression} is only slightly improved upon the Niu's model,
as manifested by its smaller MAEs and RMSEs (see Table~\ref{tab:RMSE_toughness}).
We note that the robustness and transferability of the model needs to be carefully examined due to the limited experimental data.
This should hold for the Niu's model and Oganov's model as well.

\subsection{\label{sec:level3}{Predictions of superhard materials with good fracture toughness}}

\begin{figure}
	\centering
	\includegraphics[width=0.9\textwidth]{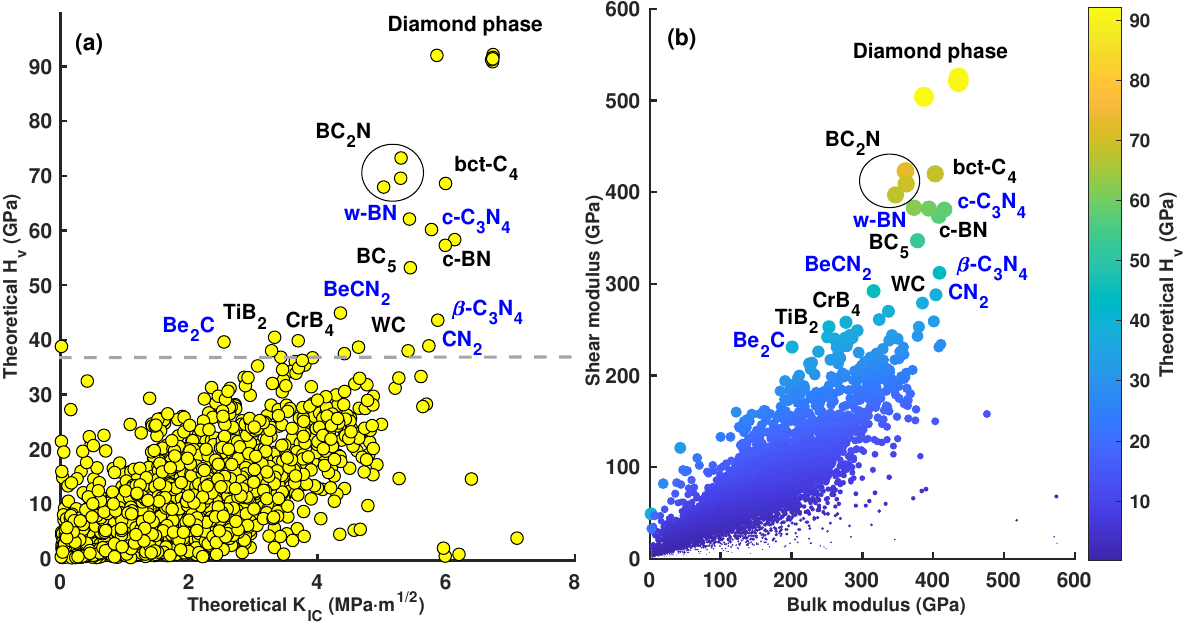}
	\caption{(a) Predicted hardness vs. fracture toughness
and (b) shear modulus vs. bulk modulus for the 8 062 materials.
The dashed line is used to identify the superhard materials (here set to 37 GPa).
The color coding in (b) indicates the predicted hardness. }
\label{fig:GBH}
\end{figure}

\begin{figure}
	\centering
	\includegraphics[width=0.9\textwidth]{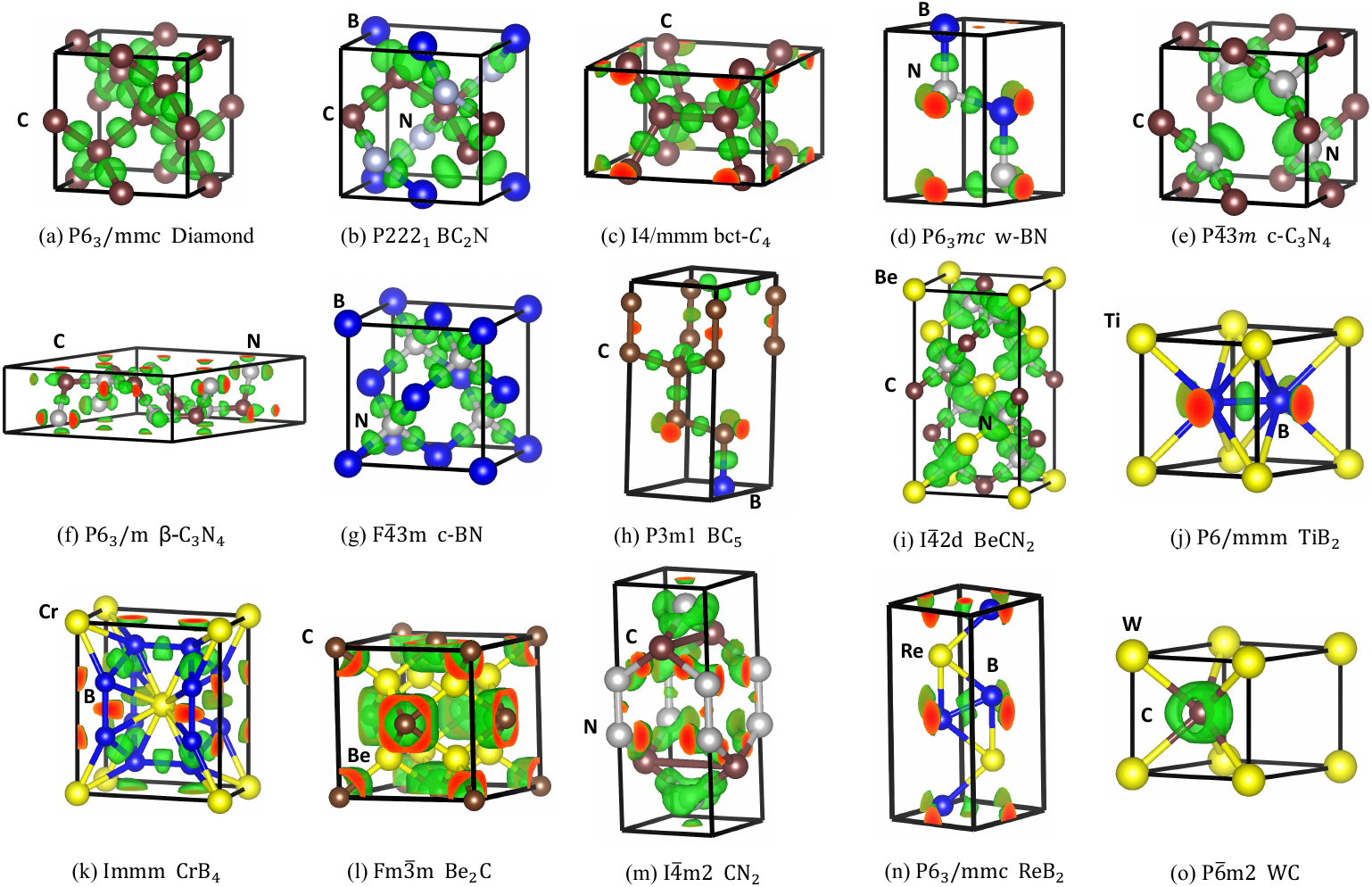}
	\caption{Isosurfaces of electron localization functions  with an isovalue of 0.80
for predicted superhard materials. The three-dimensional covalent bonding networks are evident for all materials.
}
\label{fig:ELF}
\end{figure}

With the accurate and transferable hardness model [Eq.~\eqref{eq:H_M2}] and fracture toughness model [Eq.~\eqref{eq:K_regression}]
from symbolic regression in hands, we are now in a position to perform high-throughput hardness and fracture toughness predictions.
To this aim, we download the shear and bulk moduli of 8 062 materials from the Materials Project website~\cite{articleMP}
via Python Materials Genomics (pymatgen) package~\cite{ONG2013314}.
The predicted hardness vs. fracture toughness are shown in Fig.~\ref{fig:GBH}(a).
One can see that in addition to the known superhard materials
(e.g., diamond-carbon, metastable bct-C$_4$~\cite{articleC4,articlebct-C4}, BCN$_2$~\cite{articleBeCN2PRB,articleBeCN2}, CrB$_4$~\cite{articleCrB4}, etc.)
used for training the model,
several new potentially superhard materials (marked in blue color) have been predicted.
Interestingly, most of them belong to carbon nitrides, carbides, and transition-metal borides.
For instance, C$_3$N$_4$~\cite{articleC3N4} and Be$_2$C~\cite{KALARASSE20081775} have been theoretically predicted to be superhard.
The isosurfaces of electron localization functions for these materials are shown in Fig.~\ref{fig:ELF},
exhibiting a common feature of three-dimensional covalent bonding networks.
Among these superhard materials, many of them exhibit good fracture toughness as well,
such as diamond-carbon, bct-C$_4$, BC$_2$N, c-C$_3$N$_4$, c-BN, and so on.
It should be noted that the fracture toughness model [Eq.~\eqref{eq:K_regression}]
only applies to covalent and ionic crystals, and would fail for metals and alloys.
However, this is not a big problem here, since we are only interested in
the superhard regime where most of materials are covalent and ionic crystals.
Finally, it is instructive to plot the predicted hardness as a function of $G$ and $B$.
This is done in Fig.~\ref{fig:GBH}(b) with the color coding indicating the predicted hardness.
One can see that there is a strong correlation between the shear modulus and bulk modulus.
This might explain why the simplest Teter's model~\cite{teter_1998} involving
only $G$ also works remarkably well (see Table~\ref{tab:RMSE_H}).

\subsection{Machine learning  bulk and shear moduli}\label{sec:level4}

As mentioned before, the hardness model [Eq.~\eqref{eq:H_M2}] and fracture toughness model [Eq.~\eqref{eq:K_regression}]
only require two input quantities, namely, $G$ and $B$.
Although they can be routinely calculated from first-principles, the calculations
for a large mount of materials still represent a big computational effort.
By contrast, machine learning methods
run many orders of magnitude faster than DFT calculations,
and allow one to efficiently establish the relationship between properties and atomic structure
with high precision~\cite{Xie2017CrystalGC,Avery2019PredictingSM,articleAlignn,
articleCrysXPP,Mazhnik2020ApplicationOM,articleMegnet}.

So far, several machine learning methods have been developed to predict properties of
crystalline and molecular materials simply from the structural information.
The successful models include, e.g.,
crystal graph convolutional neural network (CGCNN)~\cite{Xie2017CrystalGC},
atomistic line graph neural network (ALIGNN)~\cite{articleAlignn},
and crystal explainable property predictor (CrysXPP)~\cite{articleCrysXPP}.
From the perspective of building descriptors, CGCNN, and CrysXPP primarily reply on atomic distances,
while ALIGNN captures the many-body interactions using GCNN.
These methods differ also in input data processing and working paradigm.
Specifically,
CGCNN generates crystal graphs from crystal materials and builds a graph convolution based supervised model~\cite{Xie2017CrystalGC}.
ALIGNN presents a GNN architecture that performs message passing on both the interatomic bond graph and its line graph corresponding to bond angles,
leading to improved performance on multiple atomistic prediction tasks~\cite{articleAlignn}.
CrysXPP first learns large amounts of crystal data through an autoencoder (CrysAE),
and the learned information is then used to build a property predictor (CrysXPP)~\cite{articleCrysXPP}.

\begin{figure}
	\centering
	\includegraphics[width=0.93\textwidth]{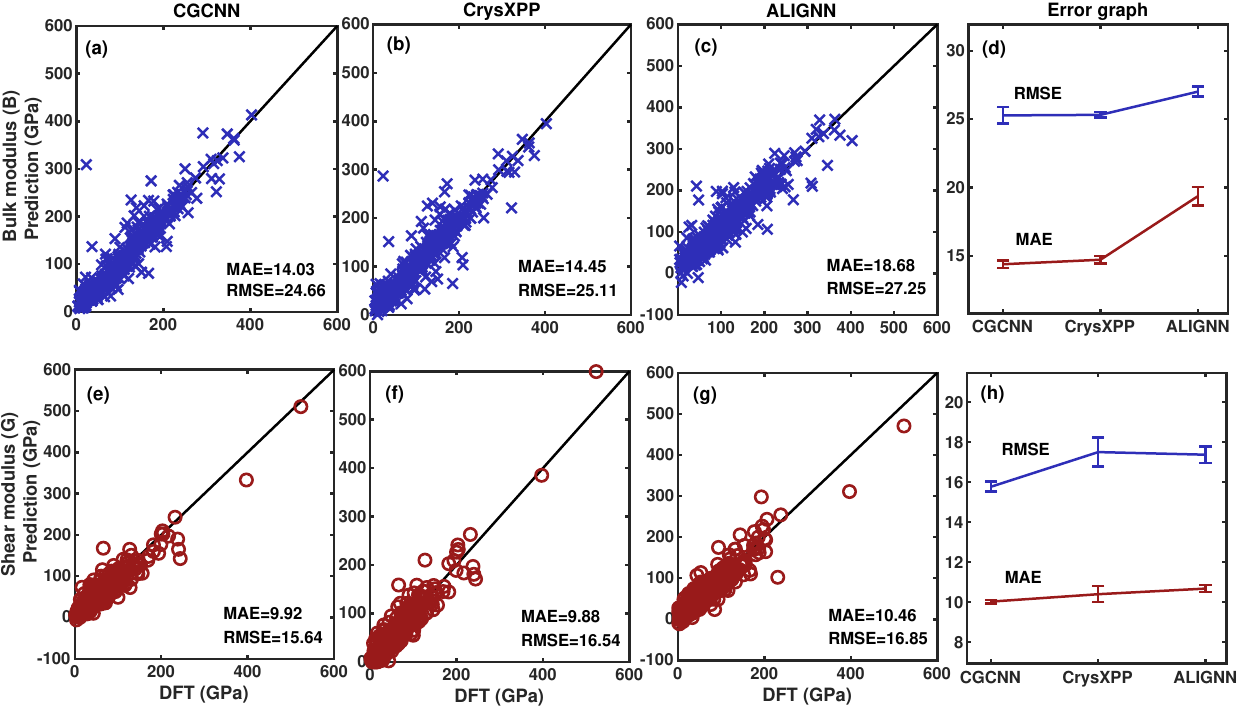}
	\caption{(d)-(h) The validation MAEs and RMSEs for the three considered deep
learning models (CGCNN, CrysXPP and ALIGNN) in predicting $B$ and $G$
with respect to the ground truth from DFT calculations.
The predicted $B$ ($G$) vs. DFT calculated ones for each model
with the lowest errors  are shown in (a)-(g).
}
\label{fig:ML_GB}
\end{figure}

Here, we employed these three advanced deep learning models to predict $G$ and $B$ of materials and assessed their performance.
The $G$ and $B$ data of  8 062 materials were taken from the Materials Project~\cite{articleMP}.
In order to make a fair assessment on the performance of different models,
the same 7 562 structural data were used for training, while the remaining 500 data were used as a test dataset.
For the CrysXPP, the structure and energy properties were learned through CrysAE using more than 30 000 data from the Materials Project~\cite{articleMP}.
Note that the 500 structures in the test dataset were not used in this step.
Then, the $G$ and $B$ of 7 562 structures were used to build CrysSPP.
For each considered deep learning models, four independent trainings were performed.
Figs.~\ref{fig:ML_GB}(d) and (h) display the validation MAEs and RMSEs of the three models
in predicting $B$ and $G$, respectively.
One can observe that the CGCNN seems to perform best among the three models.
The CrysXPP performs almost equally well as the CGCNN,
while the ALIGNN is the worst with the largest validation errors for both $B$ and $G$.
Moreover, we found that the convergence of ALIGNN is much slower than the other two models,
making the CGCNN and CrysXPP stand out for efficient and accurate prediction of shear and bulk moduli.

\section{\label{sec:level4}{Conclusions}}

In conclusion, we have built an improved macroscopic hardness model and fracture toughness model
by means of an unbiased symbolic regression technique.
The obtained models are simple, accurate, and transferable.
This is achieved by extending the existing training dataset, incorporating also materials with anisotropic elastic properties,
and comparing the widely used semiempirical Chen's hardness and Niu's fracture toughness models.
Furthermore, we have assessed deep learning models for
efficient and accurate prediction of shear and bulk moduli.
The developed hardness and fracture toughness predictions models
combined with efficient ML-learned bulk and shear moduli
allow us to fast screen out potential superhard materials with
good fracture toughness through high-throughput calculations.

\section{Acknowledgments}

This work was supported by the National Key R\&D Program of China (Grant No.~2021YFB3501503),
the National Natural Science Foundation of China  (Grant No.~52201030, Grant No.~52188101),
the National Science Fund for Distinguished Young Scholars (No. 51725103),
and Chinese Academy of Sciences (No. ZDRW-CN-2021-2-5).
All calculations were performed on the high performance computational cluster at the Shenyang National University Science and Technology Park.

\newpage

\appendix*
\setcounter{figure}{0} \renewcommand{\thefigure}{\arabic{figure}}
\renewcommand{\thefigure}{A\arabic{figure}}
\setcounter{figure}{0}
\setcounter{table}{0} \renewcommand{\thetable}{\arabic{table}}
\renewcommand{\thetable}{A\arabic{table}}
\setcounter{table}{0}

\section{Summary of experimental data on hardness and fracture toughness}

\setlength{\tabcolsep}{1pt}
\setlength{\LTcapwidth}{\textwidth}
\renewcommand\arraystretch{1.1}
\footnotesize
\begin{longtable*}{cccccccc}
\caption{A summary of material name, shear modules ($G$, in GPa), bulk modules ($B$, in GPa), experimental hardness ($H^{\rm exp}_V$, in GPa),
predicted hardness using the Chen's model ($H^{\rm Chen}_V$, in GPa) and the model in Eq.~\eqref{eq:H_M2} ($H^{\rm R2}_V$, in GPa),
and structure type for the dataset $T^A$. This is the same dataset as was used in the Chen's work~\cite{CHEN20111275}, most of which are cubic structures.
}
\label{tab:TA} \\
	\hline
	\hline
& Material & ~~~~~~$G$~~~~~~ & ~~~~~~$B$~~~~~~ & ~~$H^{\rm exp}_V$~~& ~~$H^{\rm Chen}_V$~~ & ~~$H^{\rm R2}_V$~~  & Structure \\
    \hline
    \endfirsthead
    \multicolumn{8}{l}{(Continued)} \\
    \hline
& Material & ~~~~~~$G$~~~~~~ & ~~~~~~$B$~~~~~~ & ~~$H^{\rm exp}_V$~~& ~~$H^{\rm Chen}_V$~~ & ~~$H^{\rm R2}_V$~~  & Structure \\
    \hline
      \endhead
	\hline
	\hline
	\endlastfoot
1& Diamond\cite{CHEN20111275} & 535.5 & 442.3  & 96  & 95.7 & 94.3 & cubic           \\
     &  & 548.3  & 465.5    & 96 & 93.9 & 95.2 & cubic            \\
     &  & 520.3 & 431.9     & 96  & 94.4 & 91.4     & cubic   \\
     &  & 535.0 & 443.0   & 96  & 96.6   & 94.1   & cubic     \\
2& BC$_{2}$N\cite{CHEN20111275}  & 446.0 & 403.0   & 76  & 76.9   & 75.1     & cubic   \\
     &  & 445.0 & 408.0    & 76  & 75.4    & 74.4     & cubic    \\
 & BC$_{2}$N\cite{JIANG20112287}  & 414.0  & 400.0   & 76    & 67.7    & 67.4     & cubic  \\
  &  & 414.0  & 370.0      & 76   & 74.5   & 70.1     & cubic       \\
3  & BC$_{5}$\cite{CHEN20111275}  & 394.0  & 376.0   & 71     & 66.7     & 64.5  &cubic  \\
4  & c-BN\cite{CHEN20111275}   & 405.4  & 400.0   & 66  & 65.1     & 65.3     & cubic   \\
     &  & 403.4   & 403.7    & 66    & 63.8     & 64.5     & cubic    \\
     &  & 382.2   & 375.7     & 66   & 63.1    & 61.7     & cubic            \\
     &  & 404.4    & 384.0     & 66   & 68.2  & 66.4     & cubic           \\
5 & BN \cite{JIANG20112287} & 409.0  & 400.0   & 63  & 66.2    & 66.2   & cubic        \\
6 & $\gamma$-B$_{28}$\cite{CHEN20111275}  & 236.0 & 224.0  & 50  & 49.0 & 38.8 & cubic  \\
7 & B$_{6}$O\cite{CHEN20111275}  & 204.0  & 228.0 & 38 & 36.4   & 30.9   & trigonal\\
     &  & 204.0  & 228.0   & 35  & 36.4  & 30.9     & trigonal \\
8 & $\beta$-SiC\cite{CHEN20111275}   & 191.4  & 224.7& 34 & 32.8   & 28.3 & cubic  \\
     &  & 196.6   & 224.9     & 34    & 34.5    & 29.4     & cubic            \\
     &  & 190.2  & 209.2     & 34     & 35.5    & 29.0     & cubic           \\
     &  & 186.5   & 220.3     & 34    & 32.1   & 27.5     & cubic           \\
9  & SiC\cite{JIANG20112287}  & 196.0   & 226.0    & 26   & 34.1  & 29.2     & cubic  \\
10 & SiO$_{2}$\cite{CHEN20111275} & 220.0  & 305.0    & 33   & 29.0 & 29.9 & cubic \\
   & SiO$_{2}$\cite{Mazhnik2019AMO}  & 233.1   & 324.1  & 33  & 30.0   & 31.6 & cubic\\
11 & ReB$_{2}$\cite{CHEN20111275}  & 273.0   & 382.0  & 30    & 32.9   & 36.9& hexagonal
 \\
   & ReB$_{2}$\cite{https://doi.org/10.1002/adma.200703025},\cite{osti_1419889} & 304.2  & 360.0  & 26.6  & 43.6   & 44.7     & hexagonal \\
      & & 283.0 & 350.0  & 26.6  & 39.4 & 40.7     & hexagonal   \\
12  & WC\cite{CHEN20111275} & 301.8 & 438.9   & 30   & 33.4  & 40.0& hexagonal \\
       &   & 282.0    & 439.0    & 30   & 29.3   & 36.2     & hexagonal  \\
13 & B$_{4}$C\cite{CHEN20111275}   & 192.0 & 226.0& 30 & 32.8  & 28.3 & trigonal
   \\
   & B$_{4}$C\cite{JIANG20112287}  & 171.0  & 247.0 & 30 & 23.3   & 22.8 & trigonal\\
14 & VC\cite{CHEN20111275} & 209.1  & 305.5  & 29  & 26.2   & 27.7 & cubic    \\
15 & VC$_{0.88}$\cite{JIANG20112287}  & 160.0  & 398.0  & 29 & 10.4 & 16.2   & cubic   \\
16  & ZrC\cite{CHEN20111275} & 169.7 & 223.1  & 25.8  & 26.3 & 23.7 & cubic    \\
        &  & 182.5 & 228.3       & 25.8  & 29.4   & 26.1     & cubic            \\
        &  & 185.9& 228.0    & 25.8  & 30.5  & 26.9     & cubic            \\
        &  & 169.6& 223.3      & 25.8   & 26.2    & 23.6   & cubic    \\
         &  & 166.0  & 223.0     & 25.8  & 25.2    & 22.9     & cubic           \\
     & ZrC\cite{JIANG20112287}   & 160.0  & 223.3    & 25.8& 23.4    & 21.7 & cubic  \\
17 & TiC\cite{CHEN20111275}   & 182.2  & 242.0   & 24.7  & 27.1    & 25.3     & cubic    \\
          &  & 176.9 & 250.3     & 24.7  & 24.5  & 23.8  & cubic     \\
         &  & 198.3   & 286.0  & 24.7   & 25.8  & 26.4     & cubic     \\
         &    & 187.8  & 241.7   & 24.7    & 28.8    & 26.5 & cubic     \\
      & TiC\cite{JIANG20112287}  & 188.0  & 241.0   & 28.5   & 29.0   & 26.6& cubic   \\
18 & TiN\cite{CHEN20111275}  & 183.2   & 282.0    & 23 & 22.4  & 23.6  & cubic   \\
       &  & 187.2  & 318.3  & 23  & 19.9  & 23.0  & cubic \\
     &    & 205.8  & 294.6   & 23  & 26.7   & 27.5     & cubic    \\
      & & 207.9   & 326.3      & 23   & 23.8  & 26.6     & cubic            \\
    & TiN\cite{JIANG20112287}  & 160.0  & 292.0    & 20   & 16.3    & 18.9 & cubic   \\
19   & RuO$_{2}$\cite{CHEN20111275}  & 142.2  & 251.3    & 20 & 15.7 & 17.1     & cubic   \\
           & & 173.0  & 248.0      & 20 & 23.7     & 23.1     & cubic            \\
20   & AlO$_{2}$\cite{CHEN20111275}  & 161.0  & 240.0  & 20    & 21.5   & 21.1 &orthorhombic \\
             &  & 160.0 & 259.0     & 20   & 19.2    & 20.1     & orthorhombic\\
          &    & 164.0   & 254.0     & 20    & 20.7    & 21.1     & orthorhombic\\
           & & 162.0   & 246.0        & 20   & 21.1   & 21.0     & orthorhombic\\
21   & NbC\cite{CHEN20111275}  & 171.0  & 333.0     & 18    & 15.6    & 19.6 & cubic  \\
                &  & 171.7   & 340.0    & 18  & 15.2   & 19.5     & cubic     \\
     & NbC\cite{JIANG20112287}  & 150.0 & 340.0   & 18.8  & 11.4  & 15.9  & cubic \\
22   & AlN\cite{CHEN20111275}  & 134.7 & 206.0    & 18   & 18.4    & 17.4 & cubic  \\
              && 130.2    & 212.1        & 18  & 16.5  & 16.3     & cubic   \\
                         &  & 123.3  & 207.5   & 18   & 15.2   & 15.2     & cubic    \\
        &    & 132.0  & 211.1        & 18  & 17.1 & 16.7     & cubic     \\
            & & 128.0   & 203.0          & 18   & 16.9 & 16.3     & cubic     \\
     & AlN\cite{JIANG20112287}  & 114.8  & 202.0   & 18  & 13.6   & 13.8 & cubic \\
23  & NbN\cite{CHEN20111275}  & 155.9  & 292.0    & 17  & 15.4   & 18.2     & cubic  \\
         &  & 156.0   & 315.0      & 17   & 13.9  & 17.6     & cubic    \\
    & NbN\cite{JIANG20112287}  & 165.0  & 292.0   & 20   & 17.3   & 19.8     & cubic \\
24  & HfN\cite{CHEN20111275} & 186.3   & 315.5    & 17  & 20.0& 22.9  & cubic   \\
     &  & 164.8   & 278.7      & 17  & 18.4     & 20.3     & cubic    \\
     & HfN\cite{JIANG20112287} & 202.0  & 306.0   & 19.5    & 24.5   & 26.3& cubic     \\
25   & GaN\cite{CHEN20111275}  & 105.2  & 175.9     & 15.1  & 13.7  & 13.0 & cubic   \\
         &  & 120.0   & 210.0        & 15.1  & 14.1     & 14.5     & cubic   \\
       & GaN\cite{JIANG20112287}   & 123.7  & 160.0   & 20  & 21.8   & 17.4 & cubic       \\
26   & ZrO$_{2}$\cite{CHEN20111275} & 88.0  & 187.0    & 13  & 8.4   & 9.7  & cubic  \\
             && 93.0    & 187.0         & 13  & 9.5   & 10.5     & cubic   \\
       & ZrO$_{2}$\cite{JIANG20112287}   & 80.1  & 185.0  & 13  & 6.8  & 8.4 & cubic    \\
27  & Si\cite{CHEN20111275}  & 66.6  & 97.9  & 12   & 11.9    & 8.8      & cubic    \\
       &  & 64.0   & 97.9       & 12  & 10.9    & 8.3      & cubic     \\
        &  & 63.2   & 90.7       & 12    & 11.8   & 8.4      & cubic    \\
        &  & 61.7    & 96.3       & 12   & 10.2   & 7.9      & cubic     \\
      & & 61.7   & 89.0         & 12    & 11.5    & 8.2      & cubic           \\
     & Si\cite{JIANG20112287}   & 68.0   & 98.0   & 11.3 & 12.4  & 9.1&cubic   \\
28   & GaP\cite{CHEN20111275}  & 55.7   & 88.2  & 9.5  & 9.3   & 7.1  & cubic    \\
        &   & 55.8 & 88.8         & 9.5   & 9.2   & 7.1      & cubic   \\
      &     & 56.1  & 88.6         & 9.5 & 9.4      & 7.1      & cubic    \\
        &   & 61.9  & 89.7         & 9.5  & 11.5      & 8.2      & cubic            \\
29   & AlP\cite{CHEN20111275}  & 49.0   & 86.0     & 9.4   & 7.1      & 5.9      & cubic    \\
            && 51.8    & 90.0      & 9.4  & 7.5  & 6.3      & cubic            \\
              &  & 48.8    & 86.0         & 9.4    & 7.0  & 5.9      & cubic   \\
       & AlP\cite{Mazhnik2019AMO}   & 46.8    & 81.9   & 6.5  & 6.9  & 5.7 & cubic    \\
30   & InN\cite{CHEN20111275}  & 55.0  & 123.9   & 9  & 5.1  & 5.9  & cubic   \\
            & & 77.0   & 139.6     & 9& 9.7   & 9.1      & cubic     \\
31   & Ge\cite{CHEN20111275}  & 53.1    & 72.2         & 8.8 & 11.3  & 7.3& cubic    \\
          &    & 43.8  & 60.3      & 8.8    & 9.6  & 6.0      & cubic      \\
   & Ge\cite{JIANG20112287}  & 57.0  & 71.0     & 8.8   & 13.5   & 8.2      & cubic     \\
32 & GaAs\cite{CHEN20111275}  & 46.5  & 75.0      & 7.5   & 7.8  & 5.9      & cubic   \\
          &  & 46.7   & 75.5        & 7.5  & 7.8   & 5.9  & cubic    \\
         &   & 46.7   & 75.4         & 7.5 & 7.8    & 5.9      & cubic     \\
33  & YO$_{2}$\cite{CHEN20111275}  & 72.5 & 166.0    & 7.5    & 6.3  & 7.7 & monoclinic\\
        &  & 62.7   & 146.5        & 7.5  & 5.3       & 6.6      & monoclinic\\
          &   & 66.5  & 149.3        & 7.5  & 6.0        & 7.1   & monoclinic\\
34  & InP\cite{CHEN20111275}  & 34.3 & 71.1  & 5.4  & 3.7  & 3.8 & cubic     \\
          &     & 34.4   & 72.5      & 5.4 & 3.6    & 3.8      & cubic            \\
35 & AlAs\cite{CHEN20111275}  & 44.8  & 77.9   & 5  & 6.7   & 5.4& cubic            \\
          &  & 44.6   & 78.3         & 5   & 6.5   & 5.4      & cubic    \\
36    & GaSb\cite{CHEN20111275}   & 34.2   & 56.3   & 4.5   & 5.8    & 4.3      & cubic   \\
            &   & 34.1   & 56.4       & 4.5    & 5.8   & 4.2      & cubic            \\
            &   & 34.3   & 56.3      & 4.5  & 5.9    & 4.3      & cubic       \\
37   & AlSb\cite{CHEN20111275} & 31.5  & 56.1  & 4  & 4.7    & 3.8      & cubic    \\
           &   & 31.9    & 58.2         & 4  & 4.5   & 3.8      & cubic            \\
              & & 32.5   & 59.3         & 4   & 4.6  & 3.8      & cubic            \\
              &   & 31.9    & 58.2        & 4     & 4.5   & 3.8   & cubic   \\
38   & InAs\cite{CHEN20111275} & 29.5 & 57.9   & 3.8  & 3.6   & 3.4 & cubic   \\
         &  & 29.5   & 59.1       & 3.8 & 3.4    & 3.3      & cubic    \\
39 & InSb\cite{CHEN20111275}  & 23.0  & 46.9  & 2.2  & 2.4   & 2.6& cubic    \\
         &  & 22.9   & 46.5    & 2.2   & 2.5     & 2.6   & cubic      \\
          & & 22.9    & 46.0      & 2.2  & 2.5   & 2.6      & cubic     \\
40 & ZnS\cite{CHEN20111275}  & 32.8  & 78.4    & 1.8   & 2.6       & 3.4      & cubic  \\
          &  & 31.5  & 77.1     & 1.8   & 2.3      & 3.2   & cubic      \\
41 & ZnSe\cite{CHEN20111275}  & 28.8  & 63.1   & 1.4    & 2.7   & 3.1& cubic    \\
42 & ZnSe\cite{CHEN20111275} & 23.4  & 51.0    & 1 & 2.1     & 2.5      & cubic      \\
\end{longtable*}

\setlength{\tabcolsep}{1pt}
\setlength{\LTcapwidth}{\textwidth}
\renewcommand\arraystretch{1.1}
\footnotesize
\begin{longtable*}{cccccccc}
\caption{Same as Table~\ref{tab:TA} but for the dataset $T^B$.
This dataset contains additional cubic systems that were not included in  the Chen's work~\cite{CHEN20111275}.
The values of shear and bulk moduli with asterisk are taken from the Materials Project~\cite{articleMP}.}
\label{tab:TB} \\
	\hline
	\hline
& Material & ~~~~~~$G$~~~~~~ & ~~~~~~$B$~~~~~~ & ~~$H^{\rm exp}_V$~~& ~~$H^{\rm Chen}_V$~~ & ~~$H^{\rm R2}_V$~~  & Structure \\
    \hline
    \endfirsthead
    \multicolumn{8}{l}{(Continued)} \\
    \hline
& Material & ~~~~~~$G$~~~~~~ & ~~~~~~$B$~~~~~~ & ~~$H^{\rm exp}_V$~~& ~~$H^{\rm Chen}_V$~~ & ~~$H^{\rm R2}_V$~~  & Structure \\
    \hline
      \endhead
	\hline
	\hline
	\endlastfoot
1 & $\alpha$-B\cite{JIANG20112287} & 204.5 & 224.0    & 35 & 37.4   & 31.3 & cubic  \\
2 & $\gamma-$Si$_{3}$N$_{4}$\cite{Ding_2012} &249.0* &293.0*&  35& 38.3 &36.7& cubic  \\
  &  & 249.0*   & 293.0*   & 30    & 38.7  & 36.7  & cubic  \\
3& BP\cite{JIANG20112287}  & 174.0  & 169.0    & 33 & 39.3   & 28.2 & cubic        \\
4& TaC\cite{JIANG20112287} & 190.0& 283.3      & 29  & 24.0   & 24.9     & cubic  \\
5& ZrB$_{12}$\cite{osti_1419889}  & 199.0  & 236.0    & 27   & 33.2  & 29.2  & cubic \\
6 & AlB$_{12}$\cite{JIANG20112287} & 163.0  & 139.0 & 26  & 44.4    & 28.2 & cubic  \\
7 & TaB$_{2}$\cite{JIANG20112287}  & 228.0  & 315.0   & 25.6   & 29.8 & 31.0 & cubic  \\
    & TaB$_{2}$\cite{JIANG20112287}   & 218.0   & 360.0  & 25.6  & 23.0    & 27.1 & cubic \\
8 & HfC\cite{JIANG20112287}  & 180.0  & 242.7  & 25.5  & 26.4   & 24.8  & cubic \\
& HfC\cite{Mazhnik2019AMO}& 179.2   & 238.9  & 19    & 26.7  & 24.8  & cubic   \\
9 & OsC\cite{https://doi.org/10.1002/pssb.201046127} & 72.0 & 409.0   & 20  & 0.2 & 4.8 & cubic \\
      &   & 78.0  & 438.0   & 20    & 0.4 & 5.3& cubic\\
10 & Si$_{3}$N$_{4}$\cite{JIANG20112287}  & 123.0 & 249.0   & 19  & 11.6 & 13.8  & cubic \\
11  & BAs\cite{PhysRevLett.96.085501} &124.0*& 128.0* & 19 & 29.3&19.5& cubic\\
12& Re$_{2}$C\cite{Zhao2010BulkRC} & 246.3 & 388.9  & 17.5& 26.4 & 31.4& cubic\\
13& MgB$_{2}$\cite{JIANG20112287}  & 117.5  & 145.0   & 17.4  & 22.4 & 16.9 & cubic   \\
14 & VN\cite{TIAN201293}  & 165.0* & 319.0* & 15.2& 15.3  & 19.0& cubic\\
15 & CrN\cite{TIAN201293} & 87.0*  & 181.0*  & 11& 8.6& 9.7   & cubic\\
16   & Y$_{2}$O$_{3}$\cite{Mazhnik2019AMO}  & 61.1  & 140.4  & 6.4  & 5.4 & 6.4& cubic \\
17  & MgO\cite{TIAN201293}  & 119.0*  & 151.0* & 3.9 & 21.8& 16.9 & cubic \\
18 & LiF\cite{TIAN201293} & 51.0*  & 70.0*   & 1& 10.8  & 7.0 & cubic\\
19   & CdTe\cite{Mazhnik2019AMO}   & 14.0  & 35.0   & 0.6& 0.2   & 1.4  & cubic   \\
    & & 14.0& 34.3& 0.6& 0.3& 1.4& cubic\\
20  & NaF\cite{TIAN201293} & 30.0*& 48.0* & 0.6 & 5.4 & 3.8& cubic\\
21 & NaCl\cite{PhysRevLett.96.085501}& 14.0* & 23.0*& 0.3& 2.2& 1.7 & cubic \\
22  & KCl\cite{PhysRevLett.96.085501}& 9.0* & 16.0*  & 0.2 & 0.7& 1.1& cubic \\
   & KCl\cite{TIAN201293}  & 9.0*& 16.0*  & 0.13& 0.7  & 1.1 & cubic \\
23 & KBr\cite{TIAN201293} & 7.0* & 22.0*& 0.1  & -1.4  & 0.6  & cubic\\
\end{longtable*}

\setlength{\tabcolsep}{1pt}
\setlength{\LTcapwidth}{\textwidth}
\renewcommand\arraystretch{1.1}
\footnotesize
\begin{longtable*}{cccccccc}
\caption{Same as Table~\ref{tab:TA} but for the dataset $T^C$.
This dataset contains only non-cubic systems.
The values of shear and bulk moduli with asterisk are taken from the Materials Project~\cite{articleMP}.
}
\label{tab:TC} \\
	\hline
	\hline
& Material & ~~~~~~$G$~~~~~~ & ~~~~~~$B$~~~~~~ & ~~$H^{\rm exp}_V$~~& ~~$H^{\rm Chen}_V$~~ & ~~$H^{\rm R2}_V$~~  & Structure \\
    \hline
    \endfirsthead
    \multicolumn{8}{l}{(Continued)} \\
    \hline
& Material & ~~~~~~$G$~~~~~~ & ~~~~~~$B$~~~~~~ & ~~$H^{\rm exp}_V$~~& ~~$H^{\rm Chen}_V$~~ & ~~$H^{\rm R2}_V$~~  & Structure \\
    \hline
      \endhead
	\hline
	\hline
	\endlastfoot
1 & FeB$_{4}$\cite{PhysRevLett.111.157002}& 198.0\cite{MA2019104845} & 252.0& 62&30.3& 28.1&orthorhombic \\
& & 198.0\cite{MA2019104845} & 264.7\cite{MA2019104845}  & 62 & 28.4 & 27.4 & orthorhombic\\
  & FeB$_{4}$\cite{ZHANG2017802} & 177.0& 253.0  & 62   & 24.2 & 23.7 & orthorhombic\\
      &  & 199.0  & 263.0       & 62   & 28.9  & 27.7 & orthorhombic\\
      &   & 186.0  & 277.0        & 62 & 23.7    & 24.4& orthorhombic\\
  2  & ReB$_{2}$\cite{JIANG20112287}  & 302.0& 371.0  & 48   & 41.4      & 43.6  & hexagonal   \\
3   & MnB$_{4}$\cite{ZHANG2017802} & 235.0 & 266.0   & 36   & 39.2    & 35.3 & monoclinic  \\
4  & MnB$_{4}$\cite{MA2019104845}  & 240.7 & 274.6   & 20.1  & 39.4   & 36.1  & orthorhombic \\
5 & $\beta-$Si$_{3}$N$_{4}$\cite{Gao2003HardnessOC}& 120.0*& 234.0*   & 32& 12.1   & 13.7  & hexagonal  \\
    &  & 120.0*& 234.0*  & 30  & 12.1   & 13.7& hexagonal\\
     & & 120.0*& 234.0*& 28& 12.1& 13.7& hexagonal    \\
6 & HfB$_{2}$\cite{ZHANG2017802} & 228.0  & 210.3  & 31.5 & 49.7  & 38.0& hexagonal\\
   & HfB$_{2}$\cite{doi:10.1021/acsami.1c17631}& 227.0  & 260.0 & 21.5& 37.8& 33.9& hexagonal\\
    & & 239.0& 270.0& 21.5& 39.7& 36.0& hexagonal        \\
7   & TiB$_{2}$\cite{JIANG20112287} & 263.0  & 244.0  & 31.5   & 53.9  & 43.7& hexagonal\\
      & TiB$_{2}$\cite{doi:10.1021/acsami.1c17631}  & 258.0& 250.0  & 25& 50.4& 41.9& hexagonal\\
         &  & 270.0    & 260.0  & 25    & 52.3& 44.0     & hexagonal\\
 & TiB$_{2}$\cite{osti_1419889}  & 255.0   & 240.0  & 25& 51.9& 42.1& hexagonal        \\
         & & 260.0& 250.0     & 25& 51.2& 42.4     & hexagonal       \\
8  & CrB$_{4}$\cite{ZHANG2017802}   & 296.0& 290.0& 30& 54.2& 47.8&orthorhombic\\
      &   & 259.0& 275.0& 30& 45.1& 40.2& orthorhombic     \\
    & CrB$_{4}$\cite{SIMUNEK201771}& 258.0*& 277.0*& 28.6& 44.4& 39.8&orthorhombic    \\
     &   & 258.0*& 277.0*& 23.3& 44.4& 39.8& orthorhombic     \\
9 & ZrB$_{2}$\cite{JIANG20112287}& 221.0& 218.0& 30& 44.8& 35.6     & hexagonal       \\
     & ZrB$_{2}$\cite{doi:10.1021/acsami.1c17631}& 218.0& 231.0& 17.5& 40.6& 33.9& hexagonal \\
        &   & 231.0& 228.0     & 17.5& 46.0& 37.2     & hexagonal        \\
10 & OsB$_{2}$\cite{Mazhnik2019AMO}& 167.9& 293.8& 29.4& 17.8& 20.3& orthorhombic  \\
     & OsB$_{2}$\cite{JIANG20112287} & 161.0 & 297.0 & 21.6 & 16.1  & 19.0 & orthorhombic \\
      & OsB$_{2}$\cite{SIMUNEK201771}  & 166.0*& 311.0*& 24.9& 16.1& 19.4&orthorhombic\\
  &  & 166.0*  & 311.0*  & 16.8& 16.1& 19.4     & orthorhombic     \\
    & OsB$_{2}$\cite{https://doi.org/10.1002/adma.200703025} & 166.0*  & 311.0* & 23.5  & 16.1   & 19.4& orthorhombic     \\
11   & ZrN\cite{JIANG20112287}& 160.0& 267.0& 27& 18.4& 19.8& hexagonal        \\
     & ZrN\cite{Mazhnik2019AMO}& 75.2& 196.1   & 12& 5.2& 7.4& hexagonal\\
12  & WB$_{2}$(hP6)\cite{SIMUNEK201771}& 141.0& 295.0& 27.7& 12.2& 15.6& hexagonal\\
     & WB$_{2}$\cite{osti_1419889}& 200.0& 349.0& 26.5& 20.1& 24.2& hexagonal\\
         &    & 207.0& 318.0    & 26.5& 24.4& 26.7     & hexagonal        \\
    & WB$_{2}$(hP12)\cite{SIMUNEK201771}& 224.0*& 321.0*& 22.2& 28.1& 29.9 & hexagonal\\
13  & C40-NbSi$_{2}$\cite{PU20212311}& 143.0& 175.0& 27.5& 25.8& 20.7     & hexagonal\\
14  & VB$_{2}$\cite{osti_1419889}    & 245.0& 284.0    & 25.8& 39.0& 36.4     & hexagonal\\
         &   & 241.0& 282.0   & 25.8& 38.2& 35.6     & hexagonal        \\
15  & WB$_{3}$\cite{osti_1419889}    & 249.0& 326.0& 25.5& 33.8& 34.8     & hexagonal\\
          &   & 245.0& 293.0 & 25.5& 37.5& 35.8     & hexagonal        \\
16  & MoB$_{2}$\cite{osti_1419889}   & 186.0    & 296.0 & 25.1& 21.7& 23.6& orthorhombic\\
         &   & 231.0& 303.0& 25.1& 32.1& 32.3     & orthorhombic     \\
17 & MoB$_{2}$\cite{SIMUNEK201771} & 225.0*& 295.0*& 21.2& 31.6& 31.4& trigonal\\
18 & Mo$_{2}$B\cite{SIMUNEK201771}& 141.0*& 293.0*& 24.5& 12.4& 15.7&tetragonal\\
19  & RuB$_{2}$\cite{KANOUN20101095}  & 179.2   & 289.1& 24.2     & 20.8& 22.6     & orthorhombic     \\
    & RuB$_{2}$\cite{JIANG20112287}& 144.0& 265.0 & 19.2& 14.9& 17.0& orthorhombic     \\
    & RuB$_{2}$\cite{Mazhnik2019AMO}& 183.5& 291.7& 15.1& 21.5& 23.3& orthorhombic     \\
20 & CrB$_{2}$\cite{osti_1419889}& 170.0& 295.0& 22.1& 18.2& 20.6     & tetragonal       \\
21  & CrB$_{2}$\cite{SIMUNEK201771} & 183.0*& 252.0*& 15.8& 26.0&25.0&hexagonal\\
22  & TaN\cite{TIAN201293}& 182.0*& 307.0*& 22& 19.8& 22.4& hexagonal\\
23   & Al$_{2}$O$_{3}$\cite{JIANG20112287}& 162.0  & 246.0  & 21.5& 21.1 & 21.0& trigonal \\
        &  & 165.0 & 250.6  & 17.8  & 21.3  & 21.4  & trigonal  \\
24  & CrB\cite{SIMUNEK201771} & 223.0* & 266.0*&19.6 &35.5& 32.7&orthorhombic\\
25 & Si$_{2}$N$_{2}$O\cite{Ding_2012}& 93.1 & 119.2   & 19  & 18.3  & 13.2& orthorhombic \\
26   & Mn$_{3}$B$_{4}$\cite{C8CP05870A} & 173.3& 254.0& 16.3 & 23.1  & 22.9& orthorhombic  \\
27  & BeO\cite{Mazhnik2019AMO}  & 155.0  & 206.7   & 15 & 24.3  & 21.5  & hexagonal  \\
     & BeO\cite{Mazhnik2019AMO}& 155.0& 206.7& 13\cite{Gao2003HardnessOC} & 24.3& 21.5& hexagonal\\
28  & $\alpha-$SiO$_{2}$\cite{Gao2003HardnessOC} & 44.0& 27.0& 11 & 29.4 & 9.0 & trigonal\\
29   & ZnO\cite{JIANG20112287}  & 41.0   & 148.0   & 7.2   & 0.9& 3.5   & hexagonal   \\
    & ZnO\cite{Mazhnik2019AMO} & 40.8 & 132.1   & 7.2  & 1.4  & 3.6 & hexagonal   \\
\end{longtable*}

\begin{table*}
\renewcommand\arraystretch{0.9}
\caption{The predicted fracture toughness (in MPa$\cdot$m$^{1/2}$) of covalent and ionic crystals
using the Niu's model [Eq.~\eqref{eq:Niu}] and the model in Eq.~\eqref{eq:K_regression}
as compared to the experimental data.
The data of elastic moduli ($G$ and $B$) and atomic volume ($V_0$) are taken from Ref.~\cite{Niu_2019}.
}
\begin{ruledtabular}
\begin{tabular}{cccccccccc}
&Material & G (GPa) & B (GPa)  & $V_0$ ($\AA^3/{\rm atom}$) & $K_{IC}^{\rm exp}$  & $K_{IC}^{\rm Niu}$ & $K_{IC}^{\rm R}$  \\
        \hline
		1&Diamond  & 520.3  & 431.9  & 5.70 & 5.3, 6.6, 6.7\cite{Niu_2019} &6.34&6.62\\
		2&WC& 301.8& 438.9   &  10.61& 7.5\cite{Niu_2019} & 5.40& 6.51       \\
		3&c-BN& 403.4& 403.7     & 5.95       & 5\cite{Niu_2019}  &5.43&  5.85        \\
		4&TiN& 183.2      & 282   &9.66   & 3.4, 4.28, 5.0\cite{Niu_2019}  &3.32&  3.63  \\
		5&TiC& 176.9      & 250.3   &10.19      & 2-3, 3.8\cite{Niu_2019}  &3.10&  3.23       \\
		6&$\beta-$SiC& 196.6& 224.9&10.49& 3.1, 3.3, 4.0\cite{Niu_2019} &3.11&2.99 \\
		7&Al$_2$O$_3$& 164.3& 254.3& 8.75& 3-4.5\cite{Mazhnik2019AMO} &2.93&  3.13       \\
		8&B$_4$C& 191.9& 225.8      & 7.42 & 3.08, 3.2, 3.7\cite{Niu_2019}  &2.90&  2.82    \\
		9&AlN& 122.1& 194.1       & 10.63   & 2.79\cite{Niu_2019}   &2.28&  2.30\\
		10&TiO$_2$& 110.1& 209.2     &12.22   & 2.1, 2.8\cite{Niu_2019}  &2.30&  2.47\\
		11&$\alpha$-Si$_3$N$_4$& 120.1   & 223.8  & 10.62& 3.12\cite{Niu_2019} &2.48&2.75    \\
		12&MgO& 130.3& 158.3& 9.67     & 1.9, 2.0\cite{Niu_2019}  &2.09&1.87     \\
		13&ThO$_2$& 88.1  & 187.7  &14.79 & 1.07\cite{Niu_2019}, 1.1\cite{Mazhnik2019AMO}&2.01&2.16  \\
		14&MgAl$_2$O$_4$  & 96.1 & 180.2  &9.73& 1.83, 1.94, 1.97\cite{Niu_2019}&1.92&1.98 \\
		15&Y$_2$O$_3$  & 61.3& 138.5  &15.33     & 0.71\cite{Niu_2019} &1.45&  1.46       \\
		16&ZnO$_2$   & 62.1& 113.8    &  10.15    & 1.6, 2.5\cite{Niu_2019}   &1.23& 1.13     \\
		17&Si& 66.3& 162    & 20.41& 0.79, 0.95\cite{Niu_2019} &1.33  & 1.11         \\
		18&GaP& 55.8& 88.8&21.18& 0.9\cite{Niu_2019} &1.17&0.97     \\
		19&Ge& 53.1 & 72.2&24.17 & 0.59-0.64\cite{Niu_2019}, 0.6\cite{Mazhnik2019AMO} &1.05& 0.80   \\
		20&MgF$_2$   & 52.2& 95.3&11.36    & 0.98\cite{Niu_2019}  &1.06&0.92    \\
		21&GaAs     & 46.7& 75.5      &23.92  & 0.44\cite{Niu_2019}&1.01& 0.80 \\
		22&BaTiO$_3$  & 45.1       &94.9         & 13.15      & 1.05\cite{Niu_2019}&1.01&0.91 \\
		23&InP  & 34.3    & 72.5&26.99      & 0.42-0.53\cite{Niu_2019}&0.86&0.73\\
		24&ZnS  & 32.8& 78.4  & 20.21  & 0.75, 1\cite{Niu_2019}, 0.7-1\cite{Mazhnik2019AMO}&0.84&0.74\\
		25&ZnSe   & 28.1& 58.4  &23.60       & 0.32, 1\cite{Niu_2019} &0.69&0.55\\
		26&CdS    & 18.6       & 61.1    &26.07      & 0.33, 0.76\cite{Niu_2019} &0.58&0.52\\
		27&CdSe    & 16.3       & 53.1   &29.79& 0.33-1.2\cite{Niu_2019}&0.52&0.45\\
		28&NaCl& 14.8& 24.9 &22.61  & 0.17-0.22\cite{Niu_2019}, 0.2\cite{Mazhnik2019AMO} &0.32&0.20\\
\end{tabular}
\end{ruledtabular}
\label{tab:fracture_toughness}
\end{table*}

\newpage
\bibliography{references}
\end{document}